\documentclass[12pt]{article}
\usepackage{geometry}                
\usepackage{graphicx}
\usepackage{amssymb}
\usepackage{epstopdf}
\def\bnabla{\bar\nabla}
\def\dbar{\bar D}

\def\calg{{\mathcal G}}

\def\call{{\mathcal L}}

\def\8pig{8\pi G}

\begin{document}

\begin{titlepage}
\vfill
\begin{flushright}
\end{flushright}

\vfill
\begin{center}
\baselineskip=16pt
{\Large\bf Kerr-Schild Ansatz in Lovelock Gravity}
\vskip 0.15in
\vskip 0.5cm
{\large {\sl }}
\vskip 10.mm
{\bf 
Benjamin Ett and David Kastor  \\

\vskip 1cm
{
	Department of Physics, University of Massachusetts, Amherst, MA 01003\\	
	\texttt{bett@physics.umass.edu, kastor@physics.umass.edu}
     }}
\vspace{6pt}
\end{center}
\vskip 0.2in
\par
\begin{center}
{\bf Abstract}
 \end{center}
\begin{quote}
We analyze the field equations of Lovelock gravity for the Kerr-Schild metric ansatz, $g_{ab}=\bar g_{ab} +\lambda k_ak_b$, with background metric $\bar g_{ab}$, background null vector $k^a$ and free parameter $\lambda$.  Focusing initially on the Gauss-Bonnet case, we find a simple extension of the Einstein gravity results only in theories having a unique constant curvature vacuum.   The field equations then reduce to a single equation at order $\lambda^2$.  More general Gauss-Bonnet theories having two distinct vacua yield a pair of equations, at orders $\lambda$ and $\lambda^2$ that are not obviously compatible.   Our results for higher order Lovelock theories are less complete, but lead us to expect a similar conclusion.  Namely, the field equations for Kerr-Schild metrics will reduce to a single equation of order $\lambda^p$ for unique vacuum theories of order $p$ in the curvature, while non-unique vacuum theories give rise to a set of potentially incompatible equations at orders $\lambda^n$ with $1\le n \le p$. An examination of known static black hole solutions also supports this conclusion.
  \vfill
\vskip 2.mm
\end{quote}
\end{titlepage}


\section{Introduction}

Lovelock gravities \cite{Lovelock:1971yv} are a class of higher curvature gravity models that enjoy a number of desirable properties that are not shared by generic higher curvature theories.  These include field equations that depend only on the curvature tensor and not on its derivatives \cite{Lovelock:1971yv}, the existence of ghost free constant curvature vacua \cite{Boulware:1985wk} and a reasonably well behaved initial value formulation \cite{lovelock-hamiltonian}.

Lovelock gravities have been studied in a wide variety of contexts including brane-world models beginning with the work of \cite{Bostock:2003cv}
and the gauge-gravity correspondence \cite{Brigante:2007nu,Brigante:2008gz,Buchel:2009tt,Hofman:2009ug,Shu:2009ax,deBoer:2009pn,Camanho:2009vw,Buchel:2009sk,deBoer:2009gx,Camanho:2009hu}.  Their black hole solutions have been the subject of considerable interest.  Static black hole solutions were discovered, beginning with the work of \cite{Boulware:1985wk,Wheeler:1985nh,Wheeler:1985qd} (references \cite{Charmousis:2008kc,Garraffo:2008hu} review these and related developments, including the thermodynamics of Lovelock black holes).  Many investigators have also been interested in stationary solutions, but so far only with partial success.  Rotating solutions in Gauss-Bonnet gravity with asymptotically AdS boundary conditions were found in \cite{Kim:2007iw} using first order perturbation theory in the limit of small angular momentum.  The full solutions in five dimensions were studied using numerical methods in \cite{Brihaye:2008kh,Brihaye:2010wx}.

The Kerr-Schild ansatz \cite{kerr-schild,Debney:1969zz} has been an invaluable tool for finding stationary black hole solutions in Einstein gravity.  
In addition to the four dimensional Kerr solution itself \cite{Kerr:1963ud},  the general higher dimensional Myers-Perry rotating black holes \cite{Myers:1986un} and generalizations with non-vanishing cosmological constant \cite{Hawking:1998kw,Gibbons:2004uw,Gibbons:2004js} were found in this way.  

The applicability of the Kerr-Schild ansatz in five dimensional Gauss-Bonnet gravity is studied in \cite{Anabalon:2009kq}.  The authors note earlier unpublished work leading to the expectation that Kerr-Schild would not yield a general solution for rotating black holes in Gauss-Bonnet gravity.  While the results presented in \cite{Anabalon:2009kq} are consistent with this expectation,  they do show that the Kerr-Schild ansatz leads to a solution in an interesting special case, namely when the coupling constants are such that the theory admits a unique constant curvature vacuum.

Given that is has been established \cite{Anabalon:2009kq} that the Kerr-Schild ansatz does have some range of applicability in Lovelock gravity theories, it makes sense to explore in more detail what the extent and limits of this range are.  In this paper, we will carry out this task, first looking at Gauss-Bonnet gravity in arbitrary dimension and then including arbitrary higher curvature Lovelock terms as well.  Our findings imply that the  Kerr-Schild ansatz should be expected to yield the greatest simplifications of the equations of motion for unique vacuum Lovelock theories.

The remainder of the paper proceeds as follows.  In section (\ref{lovelocksection}) we present the basic formalism of Lovelock gravity theories.  In section (\ref{kerrschildsection}) we present the Kerr-Schild ansatz for the spacetime metric, providing expressions for its curvature that are tailored to its application in Lovelock gravity.  We also present the analysis of the Kerr-Schild ansatz in Einstein gravity as a model.   In section (\ref{gaussbonnetsection}) we turn to our analysis of the Kerr-Schild ansatz in Gauss-Bonnet gravity and in  section (\ref{bhsection}) review the explicit, static Gauss-Bonnet black holes solutions in light of our results.  The general Lovelock analysis is presented in section  (\ref{generallovelocksection}) and we conclude briefly in section (\ref{discussion}).

\section{Lovelock gravity}\label{lovelocksection}

The Lagrangian for Lovelock gravity \cite{Lovelock:1971yv} has the form $\call = \sum_{k=0}^{p} c_k\call_k$
where 
\begin{equation}\label{lovelock}
\mathcal{L}_k= {\frac{1}{2^k}} \sqrt{-g}\, \delta^{a_1\dots a_{k}b_1\dots b_k}_{c_1\dots c_{k}d_1\dots d_k} \, 
R_{a_1b_1} {}^{c_1d_1}\dots R_{a_kb_k} {}^{c_kd_k},
\end{equation}
and the $\delta$ symbol denotes the totally anti-symmetrized product of Kronecker delta functions.  
At zeroth order one has $\call_0=\sqrt{-g}$ and therefore $c_0$ is proportional to the cosmological constant.  
At linear order 
$\call_1=\sqrt{-g} R$ gives the Einstein term in the Lagrangian, while the term quadratic  term is
$\call_2 = \sqrt{-g}(R_{abcd}R^{abcd}-4R_{ab}R^{ab}+R^2)$.   The theory with coefficients $c_2\neq 0$ and  $c_p=0$ for $p\ge 3$ is known as Gauss-Bonnet gravity.

Each of the terms $\call_k$ in the Lovelock Lagrangian has an intriguing `quasi-topological' aspect.  In $D=2k$ dimensions the variation of $\call_k$ is a total derivative and for a compact space without boundary its volume integral is the topologically invariant Euler character.  Since $\call_k$ vanishes identically for $D<2k$, this implies that $\call_k$ makes a non-trivial contribution to the classical dynamics only for $D\ge 2k+1$.  For example, Gauss-Bonnet gravity differs from Einstein gravity only for  $D\ge 5$.  As a consequence, one can assume the maximal order $p$ in the Lovelock Lagrangian is restricted to the ranges $p < D/2$ for even dimensions and 
$p\le (D-1)/2$ for odd dimensions.  



Unlike generic higher derivative gravity theories, 
the equations of motion of Lovelock gravity theories depend only on the curvature tensor and not on its derivatives.  
%
The equations of motion for the Lovelock theory with maximal order $p$ in the curvature tensor have the form
$\calg^{(p)a}{}_b=0$, where $\calg^{(p)a}{}_b$ may be written in terms of a new set of parameters $\alpha_0,\dots,\alpha_p$ in the form
\begin{equation}  \label{productform}
\mathcal{G}^{(p)a}{}_b=\alpha_0\, \delta^{a c_1\dots c_p d_1\dots d_p}_{b e_1\dots e_p f_1\dots f_p}\,
\left(R_{c_1d_1}
{}^{e_1 f_1}-\alpha_1 \delta^{e_1 f_1}_{c_1d_1}\right) \cdots
\left(R_{c_pd_p} {}^{e_p f_p}-\alpha_p
\delta^{e_p f_p}_{c_p d_p}\right).
\end{equation}
%
The coefficients $c_k$ in the Lovelock Lagrangian are given by sums of products of the
parameters $\alpha_k$ (see \cite{Crisostomo:2000bb} for the explicit form of this relation). Inverting
this relation to get the $\alpha_k$'s in terms of the $c_k$'s
requires solving a polynomial equation of order $p$.   Given that the coefficients $c_k$ are real numbers,  the collection of 
$\alpha_k$'s will then generally include complex conjugate pairs, although the overall tensor $\calg^{(p)a}{}_b$ will be real.  For convenience in the following we will without loss of generality assume that $\alpha_0=1$.

Real values for some number of the coefficients $\alpha_k$ with $k\ge 1$ in (\ref{productform}) have important physical significance.   If {\it e.g.} $\alpha_1$ is real, then a spacetime with constant curvature vacuum $R_{ab}{}^{cd} = \alpha_1\delta_{ab}^{cd}$ will solve the equations of motion.   If the real value of $\alpha_1$ is alternatively zero, positive, or negative, then this constant curvature spacetime will be respectively a Minkowski, deSitter or anti-deSitter vacuum of the theory.  Should all the $\alpha_k$ with $k\ge 1$ be real and distinct, then the theory will have $p$ distinct constant curvature vacua.  On the other hand, if $p$ is even it is possible that all the  $\alpha_k$ with $k\ge 1$ are complex and that therefore the theory has no constant curvature solutions.  

We will be particularly interested in special cases 
that we will refer to as unique vacuum Lovelock theories for which all the $\alpha$'s are real and equal.  
These theories have been discussed in detail in reference \cite{Crisostomo:2000bb}.
As noted in the introduction, the analysis of the Kerr-Schild ansatz in these unique vacuum theories works much as it does in Einstein gravity.   Unique vacuum theories were also shown to have special properties in reference \cite{Kastor:2006vw}, where extended black brane solutions of Lovelock theories were studied.  In dimension $D=2p+1$ the unique vacuum theory with highest Lovelock interaction $\call_p$ can be rewritten as a Chern-Simons theory (see \cite{Crisostomo:2000bb}) much as Einstein gravity can in $D=3$ \cite{Achucarro:1987vz,Witten:1988hc}

\section{Kerr-Schild Ansatz in Einstein Gravity}\label{kerrschildsection}

The Kerr-Schild ansatz works by drastically reducing the complexity of the vacuum Einstein equations, yielding a set of linear equations.
Let $\bar g_{ab}$ be a solution to the vacuum Einstein equations, which we will refer to as the background metric.
The Kerr-Schild ansatz is then given by
\begin{equation}\label{ansatz}
g_{ab}=\bar g_{ab} + \lambda h_{ab},\qquad h_{ab}= k_ak_b
\end{equation}
where $\lambda$ is a real constant and the vector $k^a=\bar g^{ab}k_b$  is null with respect to the background metric.  
The key simplification comes from the fact that the inverse  metric $g^{ab}=\bar g^{ab} -\lambda h^{ab}$, where\footnote{Except where noted explicitly, from hereon indices will be raised and lowered with the background metric.} $h^{ab}=\bar g^{ac}\bar g^{bd}h_{cd}$, 
is also linear in $h_{ab}$.

The action of the covariant derivative operator for the Kerr-Schild metric on vectors can be written as 
$\nabla_a v^b = \bnabla_a v^b +C_{ac}{}^b v^c$, where $\bnabla_a$ is the covariant derivative operator for the background metric and $C_{ab}{}^c = {1\over 2}g^{cd}\left(\bnabla_a g_{bd}+\bnabla_bg_{ad}-\bnabla_dg_{ab}\right)$.
The Riemann curvature tensor $R_{abc}{}^d$ of the Kerr-Schild metric is then related to the curvature $\bar R_{abc}{}^d$ of the background metric according to 
$R_{abc}{}^d = \bar R_{abc}{}^d + \bnabla_b C_{ac}{}^d - \bnabla_a C_{bc}{}^d +C_{ac}{}^eC_{be}{}^d - C_{bc}{}^eC_{ae}{}^d$.

The real parameter $\lambda$ can be used as a formal expansion parameter.  Expansions for key quantities typically truncate after a few terms because of the simple form of the inverse metric.   For the connection coefficients one finds a two term expansion
\begin{equation}
C_{ab}{}^c = \lambda \, C_{ab}^{(1)}{}^c +\lambda^2 C_{ab}^{(2)}{}^c
\end{equation}
%
$C_{ab}^{(1)}{}^c = {1\over 2}\left(\bnabla_a k_bk^c +\bnabla_b k_ak^c-\bnabla^c k_a k_b\right)$ and 
$C_{ab}^{(2)}{}^c 
= {1\over 2} k^c\dbar k_ak_b$, where the symbol $\dbar= k^a\bnabla_a$ is the background covariant derivative along the direction of the 
null vector $k^a$.

With the Lovelock equation of motion (\ref{productform}) in mind, we consider the curvature tensor for the Kerr-Schild metric in the form $R_{ab}{}^{cd} = g^{ce}R_{abe}{}^d$, which in principle has an expansion going out to fifth order in $\lambda$.
However, computation shows the the third, fourth and fifth order terms in the expansion vanish identically, and one is left with
$R_{ab}{}^{cd} = \bar R_{ab}{}^{cd} +\lambda\, R_{ab}^{(1)}{}^{cd} + \lambda^2 \, R_{ab}^{(2)}{}^{cd}$ with
%
\begin{eqnarray}\label{R1}
R_{ab}^{(1)}{}^{cd} &=& -2\bnabla_{[a}\bnabla^{[c} k_{b]}k^{d]} +\bar R_{ab}{}^{l[c}k^{d]}k_l\\
R_{ab}^{(2)}{}^{cd} &=&k_{[a}k^{[c} A_{b]}{}^{|k|}B_k{}^{d]} +
k_{[a}(\dbar k^{[c}) A_{b]}{}^{d]}  
-k^{[c}\left[ (\dbar k_{[a})B_{b]}{}^{d]} -    2\bnabla_{[a}(k_{b]}\dbar k^{d]})\right] \label{R2}
\end{eqnarray}
where $A_a{}^b = \bnabla_a k^b +\bnabla^b k_a$ and $B_a{}^b = \bnabla_a k^b -\bnabla^b k_a$.
It is useful to note that each term in (\ref{R2}) contains at least one factor of the null vector with no derivatives acting on it.

We are now ready to see how the Kerr-Schild ansatz linearizes the Einstein equations of motion.  Our treatment will generally follow that of 
reference \cite{xanthopoulos} adapted to the Lovelock formalism.   We may think of vacuum Einstein gravity as Lovelock gravity with maximum order $p=1$ and $\alpha_1=0$.   We work with the Einstein tensor expressed\footnote{There is an overall factor of $-1/4$ difference in normalization between the Einstein tensor $G^a{}_b=R^a{}_b- Rg^a{}_b/2$ and $\calg^{(1)a}{}_b$ in (\ref{productform}) with $\alpha_0=1$.} as in equation (\ref{productform}), 
$G^a{}_b = -{1\over 4}\, \delta^{acd}_{bef}\, R_{cd}{}^{ef}$. 
It follows that for the Kerr-Schild ansatz  the Einstein tensor will have the expansion $G^a{}_b = \lambda G^{(1)a}{}_b + \lambda^2  G^{(2)a}{}_b $ with the individual terms given by $G^{(1)a}{}_b = -{1\over 4}\, \delta^{acd}_{bef}\, R^{(1)}_{cd}{}^{ef}$ and 
$G^{(2)a}{}_b = -{1\over 4}\, \delta^{acd}_{bef}\, R^{(2)}_{cd}{}^{ef}$.
 
 Following the strategy of \cite{xanthopoulos}, we consider the implications of the contracted Einstein equations contracted twice with the 
 null vector $G^a{}_bk^b k_a=0$.  Inspection shows that the quantity $G^{(2)a}{}_bk_bk^a$ vanishes identically because of anti-symmetrization over repeated factors of the null vector.
The contracted Einstein equation then reduces to vanishing of 
\begin{eqnarray}
k_ak^b G^{(1)a}{}_b &=& {1\over 2}k_ak^b\, \delta^{acd}_{bed} \left( (\bnabla_ck_d)\bnabla^ek^f+(\bnabla_ck^f)\bnabla^ek_d\right)\label{preliminary}\\ \nonumber
&=& -{1\over 2} (\dbar k_c)(\dbar k^c)
\end{eqnarray}
This is the statement that the vector $\dbar k^a$ must itself be null with respect to the background metric.  Since $\dbar k^a$ is also orthogonal to the null vector $k^a$ itself, it follows that $\dbar k^a$ must be proportional to $k^a$.
This is then the important result that in order for the Kerr-Schild ansatz to satisfy the vacuum Einstein equations, 
the null vector $k^a$ must satisfy the geodesic condition in the background metric, so that one has $\dbar k^a = \phi k^a$ for some function $\phi$.

Let us now assume that $k^a$ satisfies the geodesic condition.  It then follows that the expression (\ref{R2}) for $R_{ab}^{(2)}{}^{cd}$ reduces to
\begin{equation}
R_{ab}^{(2)}{}^{cd}=k_{[a}k^{[c}E_{b]}{}^{d]}
\end{equation}
with $E_b{}^d = A_{b}{}^{e}B_e{}^{d} -2\phi B_b{}^d$.
Note that the expression for $R_{ab}^{(2)}{}^{cd}$ now includes factors of the null vector with indices both down and up.  
With this result it is now straightforward to show that the quantity $ G^{(2)a}{}_b$ vanishes identically for $k^a$ geodesic. The vacuum Einstein equations then reduce to the requirement that $G^{(1)a}{}_b =0$
which is  linear in  $h_{ab}$.   This is then the advertised result, that for geodesic null vector the vacuum Einstein equations for the Kerr-Schild ansatz reduce to a linear equation.  

In the following sections we will seek to find analogues of these results, first in Gauss-Bonnet gravity and then in more general Lovelock theories.
Before moving on, however, please note that although we considered only vacuum Einstein gravity in this section, the analysis is essentially unchanged for non-vacuum theories if the stress-energy tensor satisfies $T_{ab}k^ak^b=0$.  In particular, this includes Einstein gravity with a non-vanishing cosmological constant, which we may think of as the most general Lovelock gravity theory including terms up to linear order in the curvature, {\it i.e.} with maximal order $p=1$ in the Lovelock Lagrangian. 

\section{Kerr-Schild ansatz in Gauss-Bonnet gravity}\label{gaussbonnetsection}

We now turn our attention to the Kerr-Schild ansatz in Gauss-Bonnet gravity, following the same sequence of steps that we have taken in the Einstein gravity case.  As we have noted, this topic was previously studied in reference \cite{Anabalon:2009kq}.  Our analysis differs in detail from that of  \cite{Anabalon:2009kq} in a number of ways.   First, we will always be assuming that the background metric solves the field equations.   This enables the order by order expansion of the equations of motion in the parameter $\lambda$ in the Kerr-Schild ansatz (\ref{ansatz}).  The analysis of  \cite{Anabalon:2009kq} allows for more general background metrics and does not make use of such an expansion.  However, as we will see the expansion turns out to highlight the difference in applicability of the Kerr-Schild ansatz between unique and distinct vacuum Lovelock theories.
The two analyses may be regarded as complementary, with results that are related and consistent but yielding somewhat different insights.  

From equation (\ref{productform}) the field equations for Gauss-Bonnet gravity have the form $\calg^{(2)a}{}_b=0$ with
\begin{equation}\label{GBtensor}
\calg^{(2)a}{}_b = \delta^{acdef}_{bghij}\left( R_{cd}{}^{gh}-\alpha_1\delta_{cd}^{gh}\right)\left( R_{ef}{}^{ij}-\alpha_2\delta_{ef}^{ij}\right).
\end{equation}
We find the closest analogue to the results in Einstein gravity in the unique vacuum case, in which $\alpha_1$ and $\alpha_2$ are both equal to a common real value $\alpha$, and we will begin by considering this case.  We will then see how our results change when $\alpha_1$ and 
$\alpha_2$ are distinct, but still both real.  We will also assume that the background metric $\bar g_{ab}$ in the Kerr-Schild ansatz is a constant curvature vacuum of the theory,
so that
$\bar R_{ab}{}^{cd} = \alpha\delta_{ab}^{cd}$.  With these assumptions, the analysis of the field equations for the Kerr-Schild ansatz is quite similar to the Einstein case.  

We can expand the Gauss-Bonnet field equations in powers of $\lambda$ by writing 
$\calg^{(2)a}{}_b = \sum_n\lambda^n \calg^{(2,n)a}{}_b$.
Given the assumptions stated above and plugging in the nonzero terms at orders $\lambda^0$, $\lambda^1$ and $\lambda^2$ in the expansion of the curvature tensor for the Kerr-Schild ansatz
one finds
contributions to $\calg^{(2)a}{}_b$  at orders $n=2,3,4$.
%
Calculation, however, shows that $\calg^{(2,4)a}{}_b$ vanishes identically and one is left with the two term expansion 
$\calg^{(2)a}{}_b= \lambda^2\calg^{(2,2)a}{}_b+ \lambda^3\calg^{(2,3)a}{}_b$.
This result parallels the Einstein case, in which the expansion for the Einstein tensor of the Kerr-Schild ansatz also had two non-trivial terms.

Continuing to follow the model analysis from section (\ref{kerrschildsection}), we next consider the contraction $ \calg^{(2)a}{}_bk_ak^b=0$ of the Gauss-Bonnet field equation with a pair of null vectors.  
Making use of the expression (\ref{R2}) for $R^{(2)}_{ab}{}^{cd}$ one can show that
the quantity $ \calg^{(2,3)a}{}_bk_ak^b$   vanishes identically due to anti-symmetrization over repeated factors of the null vector.  
It then follows that as in Einstein gravity there is a single non-trivial term in the $\lambda$ expansion for the contracted field equations.  This  is 
given by
%
\begin{eqnarray}\label{GBgeodesic}
k_a k^b\calg^{(2,2)a}{}_b &=& -24\, (\dbar k_a)(\dbar k^b)\, \delta^{acd}_{bef}\, \alpha_{cd}{}^{ef}\\
&=& 12\, (\dbar k_a)(\dbar k^b) \calg^{(1,1)a}{}_b\nonumber
\end{eqnarray}
with $\alpha_{cd}{}^{ef} \equiv (\bnabla_ck_d)\bnabla^ek^f+(\bnabla_ck^f)\bnabla^ek_d$.  For a Kerr-Schild metric to solve the vacuum Gauss-Bonnet field equations the right hand side of (\ref{GBgeodesic}) must vanish.  Recall that in the Einstein case the analogous condition, the vanishing of the right hand side of (\ref{preliminary}), was satisfied if and only if the null vector satisfied the geodesic condition $\dbar k^a=\phi k^a$.  Note that if  $k^a$ geodesic, then the right hand side of (\ref{GBgeodesic}) is proportional to the right hand side in (\ref{preliminary}) and therefore vanishes.  This establishes that the geodesic condition is at least a sufficient condition for solving the contracted Gauss-Bonnet field equations.  

It is tempting to conjecture that it is also necessary for $k^a$ to be geodesic in order for the Kerr-Schild ansatz to solve the Gauss-Bonnet field equations.  However, we have not been able to show that this is the case.   The analysis can be advanced somewhat further by introducing a null decomposition of the background metric.  Let $l^a$ be another null vector field satisfying $\bar g_{ab}k^al^b=1$.  The background metric can then be written as $\bar g_{ab}= k_al_b +l_ak_b +\gamma_{ab}$,  where $\gamma^a{}_b$ is a spatial projection operator satisfying 
$\gamma^a{}_bk^b=\gamma^a{}_bl^b=0$.  If we additionally define the spatially projected quantities $v^a=\gamma^a{}_b\dbar k^b$ and 
$\hat\alpha_{ab}{}^{cd}= \gamma_a{}^e\gamma_b{}^f\gamma^c{}_g\gamma^d{}_h\alpha_{ef}{}^{gh}$, then one can show that the vanishing of the right hand side of (\ref{GBgeodesic}) is equivalent to the condition
\begin{equation}
v_av^b\delta^{acd}_{bef}\hat\alpha_{cd}{}^{ef}=0
\end{equation}
This form has the advantage that all the quantities involved are now spatially projected and so have positive norms.  We envision that this way of expressing the condition may be useful in either demonstrating that the geodesic condition is necessary, or in geometrically characterizing a larger set of possibilities.

We will now assume that the null vector $k^a$ is geodesic and consider the full, uncontracted Gauss-Bonnet field equations.
The key result is that calculation now shows that the quantity $\calg^{(2,3)a}{}_b$ vanishes identically.  The field equations then reduce to the single equation
\begin{equation}\label{GBKS}
\calg^{(2,2)a}{}_b = \delta^{acdef}_{bghij} R^{(1)}_{cd}{}^{gh}R^{(1)}_{ef}{}^{ij}=0,
\end{equation}
which is quadratic in $h_{ab}$.  This is a close parallel to the result in Einstein gravity, where for geodesic $k^a$ one finds that $G^{(2)a}{}_b$ vanishes identically and the field equations reduce to the single equation $G^{(1)a}{}_b=0$, which is linear in $h_{ab}$.  The result here is similar, with the single remaining equation being quadratic rather than linear in $h_{ab}$.  This result is also consistent with the findings of  
\cite{Anabalon:2009kq}.

We now consider what differences arise in the more general case that the constants $\alpha_1$ and $\alpha_2$ in the Gauss-Bonnet field equation (\ref{productform}) are unequal (but still assumed to be real), so that the theory now has two distinct constant curvature vacua.  We will assume that  the background metric has constant curvature $\bar R_{ab}{}^{cd}= \alpha_1 \delta_{ab}^{cd}$ corresponding to one of these two vacua.  Much of the analysis from the unique vacuum case ($\alpha_1=\alpha_2$) carries over to this more general case.  The key modification is now an additional linear term in the expansion of $\calg^{(2)a}{}_b$.  After assuming that the null vector $k^a$ is geodesic one is then left with a pair of equations that must be satisfied, equation (\ref{GBKS}) and also the equation linear in $h_{ab}$
\begin{equation}\label{secondequation}
\calg^{(2,1)a}{}_b = 4(\alpha_1-\alpha_2)(D-3)(D-4)\, \delta^{acd}_{bef}\,  R^{(1)}_{cd}{}^{ef}=0.
\end{equation}
Note that this latter condition is simply the vanishing of the linearized Einstein tensor.   In order to have a Kerr-Schild solution of the form (\ref{ansatz}) depending on a free parameter $\lambda$, as we have required, it is then necessary to have a solution to the linearized Einstein equations that simultaneously solves equation (\ref{secondequation}).    Note also that in the analysis of reference  \cite{Anabalon:2009kq}, which does not utilize the expansion in $\lambda$, the two conditions (\ref{GBKS}) and (\ref{secondequation}) are combined.  In the next section we will see how this added complication is manifest in the known static black hole solutions of Gauss-Bonnet gravity.  These solutions have the form (\ref{ansatz}) with free parameter $\lambda$ only in the unique vacuum case.


\section{Kerr-Schild and static Gauss-Bonnet black holes}\label{bhsection}

The static black hole solutions of Gauss-Bonnet gravity have been known for some time \cite{Boulware:1985wk,Wheeler:1985nh}.  Examining their Kerr-Schild forms will help us appreciate the difference found in the previous section between how the Kerr-Schild ansatz works in the unique vacuum case and in the more general case of distinct constant curvature vacua.  The static black hole solutions in $D\ge 5$ dimensions have the form 
$ds^2 = -f dt^2 +f^{-1}dr^2 +r^2d\Omega_{D-2}^2$ with
\begin{equation}\label{metricfunction}
f= 1-{r^2\over 2}\left\{ (\alpha_1+\alpha_2)\pm \sqrt{(\alpha_1-\alpha_2)^2 +{4\sigma\over r^{D-1}}}\right\}
\end{equation}
where $\sigma$ is a constant proportional to the black hole mass.  The metric is asymptotic at spatial infinity to the vacuum with constant curvature 
$\alpha_1$ or $\alpha_2$,  depending on which sign chosen in (\ref{metricfunction}).

Now consider the Kerr-Schild construction of static black hole metrics starting from a background metric with constant curvature $\alpha$.  This is done by writing
\begin{eqnarray}\label{staticKS}
ds^2 & = & -(1-\alpha r^2)dT^2 + {dr^2\over 1-\alpha r^2} + r^2 d\Omega_{D-2}^2
+F(r) (dT + {dr\over 1-\alpha r^2})^2 \\
&=& -(1-\alpha r^2 - F)dt^2 + {dr^2\over 1-\alpha r^2 - F} +r^2d\Omega^2_{D-2},
\end{eqnarray}
where the vector $k^a$ with covariant components $k_adx^a = dT + dr/(1-\alpha r^2)$ is null with respect to the background metric and the second line is obtained from the first by transforming to a new time coordinate $t$ such that $dt = dT + dr/(1-\alpha r^2)(1-\alpha r^2-F)$.  If one takes, for example, $F=c/r^{D-3}$ then this gives the (A)dS-Schwarzschild family of spacetimes.    By taking $F=(\alpha^\prime-\alpha)r^2$ one can also express a metric with constant curvature $\alpha^\prime$ starting from the background metric with constant curvature $\alpha$.  This is an example in which the background metric is not taken to solve the field equations.  Note also that there is no free multiplicative parameter in this case and hence are not strictly speaking of Kerr-Schild form as we have defined it in (\ref{ansatz}).

The static Gauss-Bonnet black holes with metric functions (\ref{metricfunction}) can be written in Kerr-Schild form in different ways.  Taking a flat background, $\alpha=0$ in (\ref{staticKS}), one can simply take the Kerr-Schild function $F=1-f$.  However, since the flat background will only solve the equations of motion if one of $\alpha_1$ or $\alpha_2$ is zero, this does not fit within our scheme.  Alternatively, one can start with {\it e.g.} the constant curvature background metric with $\alpha=\alpha_1$.  The Gauss-Bonnet black hole with these asymptotics is then obtained by taking 
$F=1-\alpha_1 r^2 - f$.  However, note that this function $F$ is a complicated one, as is the one for the flat background.  In particular, unlike the case of (A)dS-Schwarzschild, there is no overall free multiplicative factor in $F$.  This reflects the more complicated set of equations that the Kerr-Schild null vector must satisfy in the case of two distinct vacua.

Now consider the static black holes in the unique vacuum case, which are obtained from (\ref{metricfunction}) by setting 
$\alpha_1=\alpha_2=\alpha$.  In this limit, the metric function simplifies considerably, becoming
\begin{equation}\label{uniquebhs}
f = 1-\alpha r^2 + {\lambda \over r^{{D-5\over 2}}}
\end{equation}
with $\lambda$ a free parameter.  One can note that these spacetimes, which are discussed in some detail in reference \cite{Crisostomo:2000bb}, have slower than usual fall-off at infinity.  However, the main thing to observe is that the Kerr-Schild form of these metrics is simple.  Taking the background metric to be the unique constant curvature vacuum, the Kerr-Schild representation is simply (\ref{staticKS})  with 
$F=\lambda/ r^{{D-5\over 2}}$.  The appearance of a free multiplicative parameter in $F$ indicates that these spacetimes fall within the framework of our analysis in this paper.  We take this as evidence that looking for rotating black hole solutions via the Kerr-Schild will be simplest in the unique
vacuum case.  We plan to return to the explicit search for such solutions in future work.  Given the failure to find rotating Kerr-Schild generalizations of the static black holes of general Gauss-Bonnet theories in \cite{Anabalon:2009kq}, it may be interesting to investigate generalizations of the Kerr-Schild ansatz such as the one analyzed in \cite{Ett:2010by}.

\section{Kerr-Schild Ansatz in Lovelock}\label{generallovelocksection}

Finally, we consider Lovelock theories of arbitrary maximum order $p$ in the curvature tensor.  
We assume that the spacetime dimension $D\ge 2p+1$, so that the maximum order curvature term is dynamically relevant.
We again begin by focussing on unique vacuum theories, since our experience with Gauss-Bonnet gravity leads us to expect the strongest results in these cases.  With these assumptions the field equations have the form $\calg^{(p)a}{}_b =0$ with
\begin{equation}\label{pthorder}
\calg^{(p)a}{}_b = \delta^{ac_1d_1\dots c_pd_p}_{be_1f_1\dots e_pf_p}
\left( R_{c_1d_1}{}^{e_1f_1}-\alpha_1\delta_{c_1d_1}^{e_1f_1}\right)\cdots\left( R_{c_pd_p}{}^{e_pf_p}-\alpha_p\delta_{c_pd_p}^{e_pf_p}\right)
\end{equation}
with $\alpha_1=\dots=\alpha_p=\alpha$.
If we further assume that the background spacetime in the Kerr-Schild ansatz is the constant curvature vacuum, so that the background Riemann tensor is $\bar R_{ab}{}^{cd} = \alpha\delta_{ab}^{cd}$, then the analysis proceeds much as in the Gauss-Bonnet case.  

One can expand
$\calg^{(p)a}{}_b = \sum_n\lambda^n \calg^{(p,n)a}{}_b$ by plugging in the background curvature and the nonzero terms (\ref{R1}) and (\ref{R2}) in the expansion of the curvature tensor.
One finds immediately that $\calg^{(p,n)a}{}_b=0$ for $n>p+2$ because of anti-symmetrization over repeated factors of the null vector, while the fact that the background curvature precisely cancels the factors of $\alpha\delta^{ab}_{cd}$ in (\ref{pthorder}) implies that the lowest order nonzero term is  $\calg^{(p,p)a}{}_b$.  Computation further shows that $\calg^{(p,p+2)a}{}_b=0$  as well.  What remains is then once again a two term expansion,  with nonzero contributions at orders $\lambda^p$ and $\lambda^{p+1}$ given respectively by
%
\begin{eqnarray}
 \calg^{(p,p)a}{}_b &=& \delta^{ac_1d_1\dots c_pd_p}_{be_1f_1\dots e_pf_p} \, 
 R^{(1)}_{c_1d_1}{}^{e_1f_1}\cdots R^{(1)}_{c_pd_p}{}^{e_pf_p}\\
  \calg^{(p,p+1)a}{}_b &=& p\, \delta^{ac_1d_1\dots c_pd_p}_{be_1f_1\dots e_pf_p} \, 
R^{(2)}_{c_1d_1}{}^{e_1f_1}R^{(1)}_{c_2d_2}{}^{e_2f_2}\cdots R^{(1)}_{c_pd_p}{}^{e_pf_p}
\end{eqnarray}
These results straightforwardly generalize those found in the Einstein and Gauss-Bonnet cases which correspond to maximum Lovelock orders $p=1$ and $p=2$.

We proceed by considering the field equation contracted with a pair of null vectors, $k_ak^b\calg^{(p)a}{}_b =0$.  It follows immediately that the contraction $k_ak^b\calg^{(p,p+1)a}{}_b$ vanishes identically, once again due to anti-symmetrization over multiple factors of the null vector.
For the contraction of $G^{(p,p)a}{}_b$, a result generalizing equation (\ref{GBgeodesic}) from the Gauss-Bonnet case can be shown to hold, namely 
\begin{equation}\label{general}
\calg^{(p,p)a}{}_b k_ak^b = 2p(p+1)\calg^{(p-1,p-1)a}{}_b (\dbar k_a)\dbar k^b .
\end{equation}
It then follows by induction based on the result in Einstein gravity ($p=1$) that the geodesic condition is sufficient for the vanishing of the contracted field equations.  As in the Gauss-Bonnet case, it is possible that the geodesic condition is then either necessary as well, or that more general possibilities exist that can lead to the vanishing of (\ref{general}).

We now assume that the null vector $k^a$ is geodesic.  In analogy with the Einstein and Gauss-Bonnet cases, we would like to show that the quantity $\calg^{(p,p+1)a}{}_b$ now vanishes identically.  Although we believe that this will very likely turn out to be the case, the calculation (already long in the Gauss-Bonnet case) has so far proven too cumbersome for us to bring to completion.  Should $\calg^{(p,p+1)a}{}_b$ vanish as a consequence of the geodesic condition, then one would again be left with a single $pth$ order equation, $\calg^{(p,p)a}{}_b=0$, for the quantity $h_{ab}$ in the Kerr-Schild ansatz (\ref{ansatz}) to solve in the general Lovelock unique vacuum case.
The likelihood of this outcome is supported by the form of the black hole solutions in these theories \cite{Crisostomo:2000bb} which are very similar to those in the Gauss-Bonnet case (\ref{uniquebhs}), with metric function given by $f=1-\alpha r^2 +\lambda/r^{{D-(2p+1)\over p}}$.  This again displays the overall free multiplicative parameter, characteristic of Kerr-Schild metrics in our analysis, multiplying the departure from the background metric.

Finally, we briefly consider higher order Lovelock theories with distinct constant curvature vacua.  If we assume {\it e.g.} that all the $\alpha$'s in (\ref{pthorder}) are real and distinct and that we choose the background metric to be the constant curvature vacuum having $\bar R_{ab}{}^{cd}=\alpha_1\delta_{ab}^{cd}$, then it is straightforward to show that there will be additional equations to satisfy at orders $\lambda^n$ with $n=1,\dots,p-1$.  Hence, as in the general Gauss-Bonnet case, we would not expect to find Kerr-Schild solutions with an overall multiplicative parameter.

\section{Conclusion}\label{discussion}

Our study of the Kerr-Schild ansatz in Lovelock gravity complements and extends the earlier analysis of the Gauss-Bonnet case in \cite{Anabalon:2009kq}.  We have focused on the unique vacuum case \cite{Crisostomo:2000bb} and shown with definiteness for Gauss-Bonnet gravity, and up to plausible expectations in the general case, that the full field equations reduce to a single equation that is purely of order $p$ in the quantity $h_{ab}$ in the Kerr-Schild ansatz (\ref{ansatz}).  We have shown that the Kerr-Schild ansatz in the general Gauss-Bonnet case leads to a more complicated set of equations, which appear less promising.  Finally, we have also studied how the known static black hole solutions in the unique vacuum theories fit into our framework.  We plan to continue this work by looking for rotating black hole solutions of Kerr-Schild form in these theories.

\subsection*{Acknowledgements}

This work was supported by NSF grant PHY-0555304.

\end{document}